\documentclass[conference]{IEEEtran}

\usepackage{amssymb, amsmath}
\usepackage{algorithm}
\usepackage[noend]{algpseudocode}
\usepackage{graphicx}
\usepackage{amsthm,amsmath}
\usepackage{amsfonts}
\usepackage{subcaption}
\usepackage{mathtools}
\usepackage{color}
\usepackage{lipsum}
\newtheorem{defn}{Definition}

\def\A{\mathcal{A}}
\def\B{\mathcal{B}}
\def\Q{\mathcal{Q}}
\def\T{\mathcal{T}}

\def\N{\mathcal{N}}
\def\F{\mathcal{F}}
\def\T{\mathcal{T}}
\def\CO{\mathcal{CO}}

\begin{document}

\title{An End-to-End Framework to Identify Pathogenic Social Media Accounts on Twitter}

\author{\IEEEauthorblockN{Elham Shaabani, Ashkan Sadeghi-Mobarakeh$^{\dagger}$, Hamidreza Alvari and Paulo Shakarian}
	Arizona State University\\
	Tempe, AZ\\
	Email: \{shaabani, halvari, shak\}@asu.edu, $^{\dagger}$asade004@ucr.edu} 

\maketitle

\begin{abstract}
	Pathogenic Social Media (PSM) accounts such as terrorist supporter accounts and fake news writers have the capability of spreading disinformation to viral proportions. Early detection of PSM accounts is crucial as they are likely to be key users to make malicious information ``viral''. In this paper, we adopt the causal inference framework along with graph-based metrics in order to distinguish PSMs from normal users within a short time of their activities. We propose both supervised and semi-supervised approaches without taking the network information and content into account.  Results on a real-world dataset from Twitter accentuates the advantage of our proposed frameworks. We show our approach achieves 0.28 improvement in F1 score over existing approaches with the precision of 0.90 and F1 score of 0.63.
\end{abstract}
\IEEEpeerreviewmaketitle

\section{Introduction}
``Pathogenic Social Media'' (PSM) accounts have the capability of spreading disinformation to viral proportions. These accounts including terrorist supporter accounts, water armies, and fake news writers seek to promote or degrade certain ideas through social media in order to reach their goals. In this regard,  identifying PSM accounts has found  growing important applications to countering extremism~\cite{khader2016combating,allcott2017social}, detection of water armies~\cite{chen2013battling,chen2013best}  and fake news campaigns~\cite{kang2016fake,allcott2017social}. In Twitter, many of these accounts are social bots.

\begin{table}[ht]
	\centering
	\caption{\textmd{Label propagation selection approach - number of selected users}}
	{
		\begin{tabular}{|l| c c c |}
			\hline
			\textbf{Method} & \textbf{False Pos} & \textbf{True Pos} & \textbf{Precision} 
			\\ \hline \hline
			$All\ features$ &164,012&31,131&\textbf{0.16}\\ \hline 
			$No\ content$ &357,027&63,025&0.15\\ \hline	\hline
			$\epsilon_{K\&M}$ & 9,305 & 14,176 & 0.60 
			\\ \hline
			$\epsilon_{rel}$ & 561 &  498 & 0.47 
			\\ \hline
			$\epsilon_{nb}$ & 1,101 & 1,768 & 0.62 
			\\ \hline 
			$\epsilon_{wnb}$ & 1,318  & 4,000 & \textbf{0.75}
			\\ \hline  \hline
	\end{tabular}}
	\label{tab:pro}
\end{table}

Early detection of PSM accounts are crucial as they are likely to be the key users to malicious information campaign and detecting them is thus critical to understanding and stopping such campaigns. However, this task is difficult in practice. Existing methods rely on message content~\cite{morstatter2016new}, network structure~\cite{cao2012aiding} or a combination of both~\cite{subrahmanian2016darpa,davis2016botornot,dickerson2014using}. However, the network structure is not always available. For example, the Facebook API does not make this information available without the permission of the users. 
Moreover, the use of content often necessitates the training of a new model for the previously unobserved topics. For example, PSM accounts taking part in elections in the U.S. and Europe will likely leverage different types of content. To deal with these issues, causal inference is tailored to detect PSM accounts  in~\cite{shaabani2018detecting} based on an unsupervised learning method. The authors identified PSM users in the \emph{viral cascades}, where ``viral" is defined as an order-of-magnitude increase. As viral cascades are so rare, users that cause them are suspicious accounts. This work is continued by considering a time-decay causal and proposing a causal community detection-based classification method~\cite{alvari2018early}.

In this paper, we expand on the previous work in \cite{shaabani2018detecting} and propose graph-based metrics to distinguish PSMs from normal users within a short time around their activities. Our new metrics combined with our causal ones can achieve high precision 0.90, while increasing the recall from 0.22 to 0.49. We propose supervised and semi-supervised approaches and then show our proposed methods outperform the ones in the literature. In summary, the major contributions of this paper are itemized as follows:

\begin{itemize}
	\item  We propose supervised and semi-supervised PSM detection frameworks that do not leverage network structure, cascade path information, content and user's information.
	
	\medspace
	\item We introduce graph-based framework using the cascades and propose a series of scalable metrics to identify PSM users. We apply this framework to more than 722K users and 35K cascades.
	
	\medspace
	\item We propose a deep neural network framework which  achieves AUC of 0.82. We show that our framework significantly outperforms Sentimetrix~\cite{subrahmanian2016darpa} (0.74), causality~\cite{shaabani2018detecting} (0.73), time-decay causality~\cite{alvari2018early} (0.66), and causal community detection-based classification~\cite{alvari2018early}~(0.6).
	
	\medspace
	\item  We introduce a self-training semi-supervised framework that can capture more than 29K PSM users with the precision of 0.81. We  only used 600 labeled data for training and development sets. Moreover, if a supervisor is involved in the training loop, the proposed algorithm is able to capture more than 80K PSM users. 
	
\end{itemize}

The rest of the paper is organized as follows. In Section~\ref{sec:techapp}, we describe our framework that leverages causal metrics, graph-based metrics. We present the algorithms in Section~\ref{sec:algo}. This is followed by a description of our dataset in Section~\ref{sec:isisdata}. Then we describe our implementation and discuss our results in Section~\ref{sec:res}. Finally, related work is reviewed in Section~\ref{sec:rw}.

\section{Technical Approach} \label{sec:techapp}

\subsection{Technical Preliminaries}
Throughout this paper we shall represent cascades as an ``action log" ($Actions$) of tuples where each tuple $(u,m,t) \in Actions$ corresponds with a user $u \in U$ posting message $m \in M$ at time $t \in T$, following the convention of~\cite{goyal2010learning, shaabani2018detecting}. We assume that set $M$ includes posts/repost of a certain original tweet or message. For a given message, we  only consider the first occurrence of each user. We define $Actions_m$ as a subset of $Actions$ for a specific message $m$. Formally, we define it as $Actions_m = \{(u', m', t') \in Actions \ s.t.\ m'=m\}$.

\begin{defn} \emph{\textbf{($m$-participant).}}\label{def:cascade} 
	For a given $m\in M$, user~$u$ is an \textbf{$m$-participant} if there exists $t$ such that $(u,m,t) \in Actions$.
\end{defn}

Note that the users posting tweet/retweet in the early stage of cascades are the most important ones since they play a significant role in advertising the message and making it viral. For a given $m\in M$, we say $m$-participant $i$ ``precedes" $m$-participant $j$ if there exists $t<t'$ where $(i,m,t), (j,m,t') \in Actions$. Thus, we define \textit{key users} as a set of users adopting a message in the early stage of its life span. We formally define \textit{key user} as follows:
\begin{defn}\emph{\textbf{(Key User).}}\label{def:kusers} 
	For a given message $m$, $m$-participant $i$, and $Actions_m$, we say user $i$ is a \textbf{key user} iff user $i$ precedes at least $\phi$ fraction of $m$-participants (formally: $ |Actions_m| \times \phi \le |\{j |  \exists t': (j, m, t') \in \ Actions_m\, \wedge\, t' > t \}|$, $(i, m, t) \in Actions_m$), where $\phi \in (0,1)$.
\end{defn}
The notation  $|\cdot|$ denotes the cardinality of a set. All messages are not equally important. That is, only a small portion of them gets popular. We define \textit{viral messages} as:

\begin{defn} \emph{\textbf{(Viral Messages).}}\label{def:cascade} 
	For a given threshold $\theta$, we say that a message $m \in M$ is \textbf{viral} iff $|Actions_m| \ge \theta$. We use $M_{vir}$ to denote the set of viral messages. 
\end{defn}

The Definition~\ref{def:cascade} allows us to compute the prior probability of a message (cascade) going viral as follows:
\begin{equation}
	\rho = \frac{|M_{vir}|}{|M|}
\end{equation}

We also define the probability of a cascade $m$ going viral given some user $i$ was involved as:

\begin{equation}
	p_{m | i} = \frac{|\{m \in M_{vir}\, s.t.\ i\ is\  a \  key \ user\}|}{|\{m \in M\, s.t.\ i\ is\  a \  key \ user\}|}
\end{equation}

We are also concerned with two other measures. First, the probability that two users $i$ and $j$ tweet or retweet viral post $m$ chronologically, and both are key users. In other words, these two users are making post $m$ viral.
\begin{equation}\label{eq:rule1}
	p_{i,j} = 
	\frac{
		\splitfrac{
			|\{m \in M_{vir} | \exists t,t'\, where\, t<t'\, and
		}
		{
			\, (i,m,t),(j,m,t') \in Actions\}|
		}
	}
	{
		\splitfrac{
			|\{m \in M | \exists t,t'\, where\, t<t'\ and
		}
		{
			\ (i,m,t),(j,m,t') \in Actions\}|
		}
	}
\end{equation}

Second, the probability that key user $j$ tweets/retweets viral post $m$ and user $i$ does not tweet/retweet earlier than $j$. In other words, only user $j$ is making post $m$ viral.
\begin{equation}\label{eq:rule2}
	p_{\neg{i},j} = \frac{
		\splitfrac{
			|\{m \in M_{vir} | \exists t'\, s.t.\, (j,m,t') \in Actions\, and
		}
		{
			\, \not\exists t\  where\ t <t',\ (i,m,t) \in Actions\}|
		}
	}
	{
		\splitfrac{
			|\{m \in M | \exists t'\, s.t.\, (j,m,t') \in Actions\, and
		}
		{
			\, \not\exists t\  where\ t<t', \ (i,m,t) \in Actions\}|
		}
	}
\end{equation}

Knowing the action log, we aim to find a set of pathogenic social media (PSM) accounts. These users are associated with the early stages of large information cascades and, once detected, are often deactivated by a social media firm. In the causal framework, a series of causality-based metrics for identifying PSM users is introduced.

\subsection{Causal Framework}
We adopt the causal inference framework previously introduced in~\cite{suppes1970probabilistic,kleinberg2009temporal}. We expand upon that work in two ways~\cite{shaabani2018detecting}: (1.) we adopt it to the problem of identifying PSM accounts and (2.) we extend their single causal metric to a set of metrics. Multiple causality measurements provide a stronger determination of significant causality relationships. 

For a given viral cascade, we seek to identify potential users who likely \textit{cause} the cascade viral. We first require an initial set of criteria for such a causal user.  We do this by instantiating the notion of Prima Facie causes to our particular use case below:

\begin{defn} \emph{\textbf{(Prima Facie Causal User).}}\label{def:pf} 
	A user $u$ is a prima facie causal user of cascade $m$ iff: User $u$ is a key user of $m$,  $m \in M_{vir}$, and $p_{m|u} > \rho$.
\end{defn}

For a given cascade $m$, we will often use the language \textit{prima facie causal user} to describe user $i$ is a prima facie cause for $m$ to be viral. In determining if a given prima facie causal user is causal, we must consider other ``related" users. We say $i$ and $j$ are $m$-related if (1.) $i$ and $j$ are both prima facie causal users for $m$, (2.) $i$ and $j$ are both key users for $m$, and (3.) $i$ precedes $j$. Hence, we will define the set of ``related users" for user $i$ (denoted $R(i)$) as follows:
\begin{equation}\label{eq:R}
	R(i) = \{j \ s.t. \ j \not= i\ , \exists m \in M \ s.t. \ i,j \ are \ m-related\}
\end{equation}

Therefore, $p_{i,j}$  in (\ref{eq:rule1}) is the probability that cascade $m$ goes viral given both users $i$ and $j$, and $p_{\neg i,j}$ in (\ref{eq:rule2}) is the probability that cascade $m$ goes viral given key user $j$ tweets/retweets it while key user $i$ does not tweet/retweet $m$ or precedes $j$. The idea is that if $p_{i,j} - p_{\neg i,j} > 0$, then user $i$ is more likely a cause than $j$ for $m$ to become viral. We measure \textit{Kleinberg-Mishra causality} ($\epsilon_{K\&M}$) as the average of this quantity to determine how causal a given user $i$ is as follows:
\begin{equation}
	\epsilon_{K\&M}(i) = \frac{\sum_{j\in R(i)} (p_{i,j} - p_{\neg i, j})}{|R(i)|}
\end{equation}
Intuitively, $\epsilon_{K\&M} $ measures the degree of causality exhibited by user $i$.  Additionally, we find it useful to include a few other measures. We introduce \textit{relative likelihood causality} ($\epsilon_{rel}$) as follows: 
\begin{equation}
	\epsilon_{rel}(i) = \frac{\sum_{j\in R(i)} S(i,j)}{|R(i)|}
\end{equation}
\begin{equation}
	S(i, j)=
	\begin{cases}
		(\frac{p_{i,j}}{p_{\neg i, j} + \omega})-1 , & p_{i,j}> p_{\neg i, j}\\
		0, & p_{i,j} = p_{\neg i,j'}\\
		1- (\frac{p_{\neg i, j}}{p_{i,j}}), &\text{otherwise}
	\end{cases}
\end{equation}

where $\omega$ is infinitesimal. Relative likelihood causality metric assesses the relative difference between $p_{i,j}$ and $p_{\neg{i},j}$. This helps us to find new users that may not be prioritized by $\epsilon_{K\&M}$.
We also find that if a user is mostly appearing after those with the high value of $\epsilon_{K\&M}$, then it is likely to be a PSM account. One can consider all possible combinations of events to capture this situation. However, this approach is computationally expensive. Therefore, we define $\Q(j) $ as:
\begin{equation}\label{eq:Q}
	\Q(j) = \{i \ s.t. \ \ j \in R(i)\}
\end{equation}

Accordingly, we define  \textit{neighborhood-based causality}~($\epsilon_{nb}$) as the average $\epsilon_{K\&M}(i)$ for all $i \in Q(j)$ as follows: 
\begin{equation}
	\epsilon_{nb}(j) = \frac{\sum_{i \in \Q(j)} \epsilon_{K\&M} (i)}{|\Q(j)|}
\end{equation}

The intuition behind this metric is that accounts who are retweeting a message that was tweeted/retweeted by several causal users are potential for PSM accounts. We also define the \textit{weighted neighborhood-based causality}~($\epsilon_{wnb}$) as follows:
\begin{equation} \label{eq:weighted}
	\epsilon_{wnb} (j) = \frac{\sum_{i \in \Q(j)} w_i \times \epsilon_{K\&M} (i) }{\sum_{i \in \Q(j)} w_i}
\end{equation}

The intuition behind the metric $\epsilon_{wnb}$ is that the users in $\Q$ may not have the same impact on user $j$ and thus different weights $w_i$ are assigned to each user $i$ with~$\epsilon_{K\&M}(i)$. 
In our previous work \cite{shaabani2018detecting}, we provided more details of these metrics.

\subsection{Graph-based Framework}
\begin{figure}
	\centering
	\includegraphics[width=0.7\columnwidth]{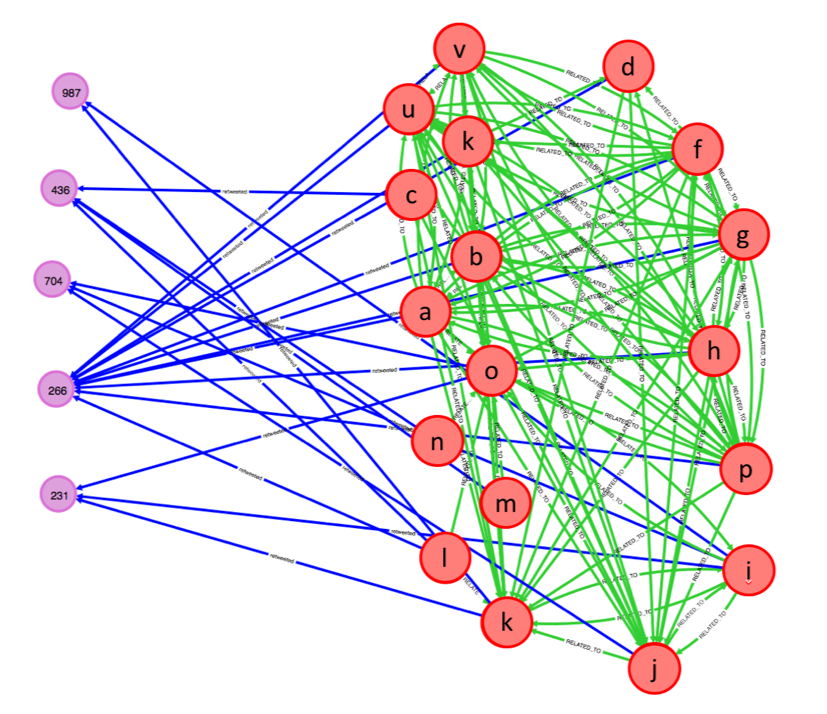}
	\caption{User-message bipartite graph and user graph. Red and purple nodes represent users and messages respectively. Messages are labeled with the length of the cascade (degree). Blue and Green edges represent user-message and user-user relationships.} 
	\label{fig:graph}	
\end{figure}

\noindent\textbf{User-Message Bipartite Graph.} Here, we denote $Actions$ as a bipartite graph $G_{u-m}(U, M, E)$, where  users $U$ and messages $M$ are disjoint sets of vertices. There is an annotated link from user $u$ to message $m$ if $u$ has tweeted/retweeted $m$ and is annotated by occurrence time $t$ (see Fig. \ref{fig:graph}). In other words, every edge in graph $G_{u-m}$ is associated with one tuple $(u,m,t) \in Actions$. For a given node $u \in U (m \in M)$, the set $\N_u= \{m' \in M\ s.t.\ (u,m') \in E\} (\N_m= \{u' \in U\ s.t.\ (u',m) \in E\} )$ is the set of immediate neighbors of $u$ ($m$). We also define  $U^v \subset U$ which is the set of verified users (often celebrities). We indicate $U^v_m = \{u|(u,m,t)\in Actions, u \in U^v\}$ as a set of verified users that have re/tweeted message $m$. 

As for the edges, we examine different metrics such as Jaccard similarity between users, and rank of a user in a message which is defined as $Rank_{(u,m)} = |\{(u',m,t')\in Actions| (u,m,t) \in Actions, t'<t \}|$.  We also define normalized rank as:
\begin{equation}
NR(u,m) = 1- \frac{Rank_{(u,m)}}{\N_m}
\end{equation}

Our intuition behind rank metric $Rank_{(u,m)}$ is that the earlier a user has participated in spreading a message, the more important the user is. In this regard, we can also define the exponential decay of the time as: 

\begin{equation}
\T_u^m = \exp (-\gamma \Delta t_u^m)
\end{equation}

where $\Delta t_u^m = \{t|(u,m,t)\in Actions\} - \min(\{t'|(u',m,t')\in Actions\})$, and $\gamma$ is a constant. This metric prioritizes based on the retweeting time of the message. In other words, this metric assigns different weights to different time points of a given time interval, inversely proportional to their duration from start of the cascade, i.e., smaller duration is associated with higher weight.

Using all these information, we then annotated users $U$ based on their local and network characteristics such as degree, and PageRank. We also consider function $ \F \in \{sum,max, min, avg, med, std\}$ to calculate statistics such as minimum, mean, median, maximum, and standard deviation based on their one-hop or two-hops neighbors. For example, for a given user $u$, mean of re/tweeted message's PageRank of user $u$ proved to be among top predictive metrics according to our experiments. Using these intuitions, we explored the space of variants features and list those we found to be best-performing in Table~\ref{tab:um_metrics}.

\begin{table}[ht]
	\centering
	\caption{\textmd{User-Message Bipartite Graph-Based Metrics}}
	\begin{tabular}{| p{3.5cm}| p {4cm}|}
		\hline
		\textbf{Name} & \textbf{Definition}   \\\hline \hline
		\vspace{0cm}	 	Degree &  \vspace*{-3mm}	\begin{equation}\nonumber
		\splitfrac{D_v = }{ |\{v'|(v,v') \in E \vee (v',v) \in E \}|}
		\end{equation} 
		 \vspace*{-3mm}
		\\ \hline
		\vspace{0cm} Cascade size statistics &  \vspace*{-1mm}\begin{equation}\nonumber
		CS_{u,\F} = \F_{m \in \N_u} D_m
		\end{equation} \vspace*{-3mm}
		\\ \hline
		\vspace{2mm}	PageRank&  \vspace*{-2mm}	\begin{equation}\nonumber
		\splitfrac{PR(v) = }{\frac{1-d}{N}+d \sum_{v' \in \N_v} \frac{PR(v')}{L(v')} }
		\end{equation}\vspace*{-2mm}\\\hline
		\vspace{0cm}	Message's PageRank statistics &\vspace*{-3mm}\begin{equation}\nonumber
		PS_{u,\F} = \F_{m \in \N_u} PR(m)
		\end{equation}\vspace*{-3mm}\\ \hline
		\vspace{0cm}	Number of verified users & \vspace*{-3mm} \begin{equation}\nonumber
		\splitfrac{Vr_m =}
		{|\{u|(u,m) \in E, u\in U^v\}| }
		\end{equation}\vspace*{-3mm}\\ \hline
		\vspace{0cm}	Jaccard similarity statistics & \vspace*{-3mm}\begin{equation}\nonumber
		JS_{u,\F} = \F_{u' \in U } \frac{|\N_u \cap \N_{u'}|}{|\N_u \cup \N_{u'}|}
		\end{equation}\vspace*{-3mm}
		\\ \hline		
		\vspace{0cm}	Intersection statistics & \vspace*{-3mm}\begin{equation}\nonumber
		IS_{u,\F} = \F_{u' \in U } |\N_u \cap \N_{u'}|
		\end{equation}\vspace*{-3mm}
		\\ \hline
		\vspace{0cm}	Normalized rank statistics & \vspace*{-3mm}\begin{equation}\nonumber
		NRS_{u,\F} = \F_{m \in \N_u} NR{(u,m)}
		\end{equation}\vspace*{-3mm}
		\\ \hline
		\vspace{0cm}	$\T$ statistics & \begin{equation}\nonumber
		\T S_{u,\F} = \F_{m \in \N_u} \T_u^m
		\end{equation}\vspace*{-3mm}\\ \hline
		\vspace{0cm}	Verified users in the
		cascades statistics &\vspace*{-2mm}  \begin{equation}\nonumber
		U^v S_{u,\F} = \F_{m \in \N_u} |U^v_m|
		\end{equation}\vspace*{-3mm}\\ \hline
		\hline
	\end{tabular}
	\label{tab:um_metrics}
\end{table}

\noindent\textbf{User Graph.}
We represent a directed weighted user graph $G(V', E')$ where the set of nodes $V'$ corresponds with key users. There is a link between two users if they are both key users of at least a message. There is a link from $i$ ($j$) to $j$ ($i$) if the number of times that ``$i$ appears before $j$ and both are key users'' is equal to or larger (smaller) than the case when ``$j$ appears before $i$'', see Fig. \ref{fig:graph}. For a given node $i$, the set $N_i^{out}= \{i' \in V'\ s.t.\ (i,i') \in E'\}$ ($N_i^{in}= \{i' \in V'\ s.t.\ (i',i) \in E'\}$)- the set of outgoing (incoming) immediate neighbors of $i$. The weight of edges is determined as a variant of co-occurences of the key user pairs:
\begin{equation}
\CO_{i,j} = \frac{
	\splitdfrac{|\{m|i,j\ are\ key\ users,\ \exists t,t'\ where\ t<t',}
	{ (i,m,t), (j,m,t') \in Actions\}|}}
{\min(|\{m|i\ is\ a\ key\ user\}|, |\{m|j\ is\ a\ key\ user\}|)}
\end{equation}

Using $\CO_{i,j}$, we then propose a weighted co-occurrence score for user $i$ as:
\begin{equation}
\CO^w_{i, N_i} = \frac{\sum_{j \in N_i} (abs(\delta_{i,j}) +1) \times \CO_{i,j} }
{\sum_{j \in N_i} (abs(\delta_{i,j}) +1)}
\end{equation}
where $abs(\cdot)$ denotes the absolute value of the input. The differences between ordered joint occurrences $\delta_{i,j}$ is also defined as:
\begin{equation}
\splitdfrac{\delta_{i,j} =}{
	\splitdfrac{ |\{m| \exists t,t'\ s.t.\ t<t', (i,m,t), (j,m,t') \in Actions\}|}
	{-|\{m| \exists t,t'\ s.t.\ t>t', (i,m,t), (j,m,t') \in Actions\}|}}
\end{equation}
The list of user graph-based metrics extracted from graph $G$ is shown in Table~\ref{tab:uu_metrics}.
\begin{table}[ht]
	\centering
	\caption{\textmd{User Graph-Based Metrics}}
	\begin{tabular}{|p{3.5cm}| p {4cm}|}
		\hline
		\textbf{Description} & \textbf{Definition}   \\\hline \hline
		\vspace{0cm} Degree & \vspace*{-1mm} \begin{equation}\nonumber
		|N_i^{out}|
		\end{equation}\vspace*{-3mm}\\ \hline
		\vspace{0cm}	Outgoing co-occurrence score statistics & \vspace*{-3mm} \begin{equation}\nonumber
		\CO S_{i,\F}^{out} = \F_{j \in \N_i^{out}} \CO_{i,j}
		\end{equation}\vspace*{-3mm}\\ \hline
		\vspace{0cm}	Incoming co-occurrence score statistics&\vspace*{-3mm} \begin{equation}\nonumber
		\CO S_{i,\F}^{in} = \F_{j \in \N_i^{in}} \CO_{i,j}
		\end{equation}\vspace*{-3mm}\\ \hline
		\vspace{0cm}	Weighted co-occurrence score & \vspace*{-1mm}\begin{equation}\nonumber
		\CO^w_{i, N_i^{out}} 
		\end{equation}\vspace*{-3mm}\\ \hline
		\vspace{0cm}	Number of outgoing verified users &\vspace*{-2mm} \begin{equation}\nonumber
		|\{j|j\in N_i^{out}, j \in U^v\}|
		\end{equation}\vspace*{-3mm}\\ \hline
		\vspace{0cm}	Number of incoming verified & \vspace*{-1mm}\begin{equation}\nonumber
		|\{j|j\in N_i^{in}, j \in U^v\}|
		\end{equation}\vspace*{-3mm}\\ \hline
		Triangles & Number of triangles\\ \hline
		\vspace{0cm}	Clustering coefficient & \vspace*{-3mm}\begin{equation}\nonumber
		\splitdfrac{CC_i = }
		{\frac{|\{(j,k) | j,k \in N_i, (j,k) \in E'  \}|}{|N_i| \times (|N_i|-1)}}
		\end{equation}\vspace*{-1mm}\\ \hline
		\hline
	\end{tabular}
	\label{tab:uu_metrics}
\end{table}
We further calculate the probability of ``user $j$ appears after user $i$'' as:
\begin{equation}
P_{(j,i)} = \frac{\splitfrac{
		|\{m \in M_{vir} | \exists t,t'\, where\, t<t'\, and
	}
	{
		\, (i,m,t),(j,m,t') \in Actions\}|
} }{|\{m|(j,m,t) \in Actions\}|}
\end{equation}
The average probability that user $i$ appears before its related users $R(i)$ is also a good indicator for identifying PSM accounts:
\begin{equation}
CM_i = \frac{\sum_{R(i)} P_{(j,i)}}{|R(i)|}
\end{equation}
We aim to evaluate users from different perspectives and these metrics have shown to be helpful for evaluating users and detecting PSM accounts.

\subsection{Problem Statement}
Our goal is to find the potential PSM accounts from the cascades. In the previous section, we discussed causality metrics, and defined diverse set of features using both user-message bipartite and user graphs where these metrics can discriminate the users of interest. 

\noindent{\textbf{Problem. (Early PSM Account Detection).}} \label{prob:1}
\emph{	Given Action log Actions, causality and structural metrics, we wish to identify set of key users that are PSM accounts.}

\section{PSM Account Detection Algorithm}\label{sec:algo}
We employ supervised, and semi-supervised approaches for detecting PSM accounts. Proposed metrics are scalable and can be calculated efficiently using map-reduce programming model and storing data in a graph-based database. To such aim, we used Neo4j to store data and calculated most of the structural metrics using Cypher query language~\cite{webber2018programmatic}.

\subsection{Supervised Learning Approach}
\begin{figure}[ht]
	\centering
	\includegraphics[width=.7\columnwidth]{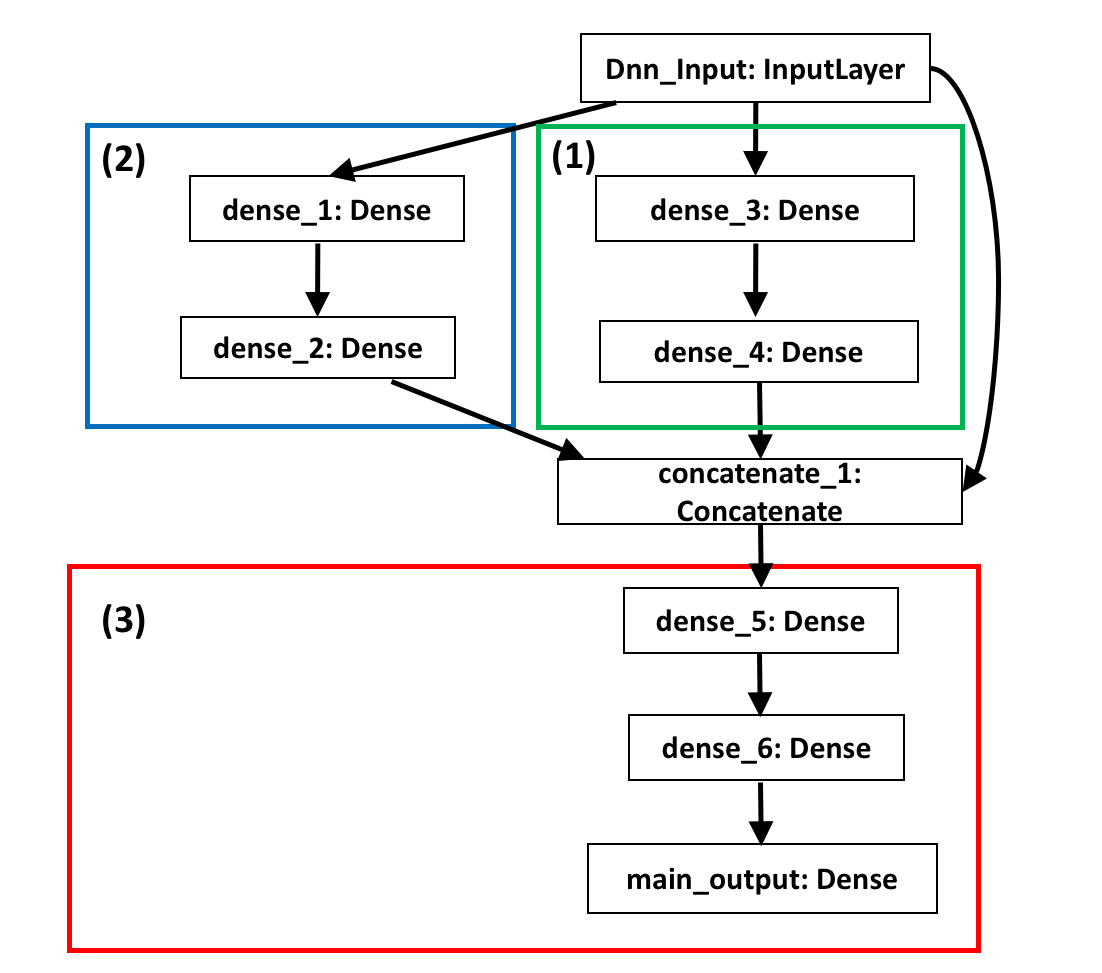}
	\caption{The proposed deep neural net structure}
	\label{fig:Dnn-Structure}
\end{figure}

We evaluate several supervised learning approaches including logistic regression (LR), naive bayes (NB), k-nearest neighbors (KNN) and random forest (RF) on the same set of features. We also develop a dense deep neural network structure using Keras. As for the deep neural network and in order to find the best architecture and hyperparamters, we utilize the random search method. Many model structures were tested and Fig.~\ref{fig:Dnn-Structure} illustrates the best architecture.

As we can see from Fig. \ref{fig:Dnn-Structure}, the proposed deep neural net, in fact, consists of three dense deep neural net structures. The first two structures are of the same, but the activation functions of their layers are different. The intuition is that we aimed to capture the most useful information from the input data and our experiments show the ReLU and Sigmoid activation functions can contribute to this.  Specifically, these two structures are aimed to filter the noises in the input data and prepare clean inputs to feed into the third  structure. In this regard, the outputs of these two structures along with the input data are   concatenated into one vector and this vector is fed into another dense deep neural net.  Finally, the output of this structure  is fed into a regular output layer.  To avoid overfitting, we used dropout method.
In the proposed framework, the binary cross entropy loss function is minimized and the best optimizers are reported as Adam and Adagrad.

\subsection{Self-Training Semi-Supervised Learning Approach}

Semi-supervised algorithms~\cite{subrahmanian2016darpa,zhu2006semi} use unlabeled data along with the labeled data to better capture the shape of the underlying data distribution and generalize better to new samples. Here, we propose a \emph{Weighted Self-Training Algorithm} (\textsc{WSeT}) shown in Algorithm~\ref{alg:WSeT} to address such problem. We start with small amount of labeled training data and iteratively add users  with high confidence scores from unlabeled data to the training set. Lets denote labeled data $L=\{\textbf{u}_i, l_i\}$ and unlabeled data $U=\{\textbf{u}_j\}$. Labeled data is split to training set $L_t$ and development set $L_d$. We then iteratively train a classifier using training set and predict the confidence scores for development set and unlabeled data.  Based on the \emph{confidence score} obtained from  \emph{development set},  a \emph{threshold} is determined. We then select \emph{all samples} from \emph{unlabaled data} that \emph{satisfy the threshold}. Next, those samples are \emph{removed}  from \emph{unlabeled set} and are \emph{added}  to the \emph{training set}. The termination condition is determined based on  at most  $\theta_{tr}$ drop in accuracy on the development set or minimum number of selected users by algorithm.

\begin{algorithm}
	\caption{\small \textbf{Weighted Self-Training Algorithm (\textsc{WSeT})}}\label{alg:WSeT}
	{
		\begin{algorithmic}[1]
			\Procedure{WSeT}{$L=\{\textbf{u}_i, l_i\}, U=\{\textbf{u}_j\}, \alpha, \beta, \theta_{pr}, \theta_{tr}$}
			\State {Split $L$ to training set $L_t$ and development set $L_d$}
			\State {$L_t.w_c = 1$}
			\State {$it = 1$}
			\State {$m = $ Train a classification model using $L_t$}
			\State {$L_d.p=$ confidence score $p$ using $m$ of $L_d$}
			\State {$c =$ accuracy of model $m$ on $L_d$}
			\State{$c'=c$}
			\While {$c'>= c-\theta_{tr}$}
			\State {$U.p=$ confidence score $p$ using $m$ of $U$}
			\State {Update $L_t$ and $U$ by Algorithm~\ref{alg:UpDWSeT} ($L_t, L_d, U, \alpha,\beta, \theta_{pr}, it$) }
			\State {$m= $ Train a classification model using $L_t$}
			\State {$L_d.p=$ confidence score $p$ using $m$ of $L_d$}
			\State {$c' =$ accuracy of model $m$ on $L_d$}
			\State {$it = it + 1$}
			\EndWhile
			\State \textbf{return} $L_t$
			\EndProcedure
	\end{algorithmic}}
\end{algorithm}

There are still two main questions that need to be answered:
\begin{description}
\item[Q1.]   Should all training samples be weighted equally?
\item[Q2.] How should a threshold be determined for adding unlabeled data to the labeled set? 
\end{description}

Since the prediction mistake reinforces itself, and the prediction error increases by number of iterations, the way we choose samples is of importance. According to our experiments, all training samples should not be weighted equally. We found the \emph{exponential decay weighting} approach as the most efficient one (see Q1). Considering a sample  with confidence $p_l$ associated to a specific label $l$ in iteration $it$, the exponential decay weighting approach  is  defined as:

\begin{equation}
\exp(-\beta \times it \times (\frac{1}{1-p_l}))
\end{equation}
where $\beta$ is a parameter. To answer the second question, we pick the threshold to have the minimum precision of $\theta_{pr}$ on development set in each iteration. Since the precision decreases as the algorithm iterates, the threshold is required to be adjusted in order to make sure  the top ranked and qualified samples are picked up. Mathematically, the updated threshold  in each iteration is defined as follows:

\begin{equation}
\theta_{pr}-\alpha \times (it-1)
\end{equation}
where $\alpha$ is a parameter, $\alpha \in [0, \frac{1}{it-1}], it>0$. We pick $0.005$ for the experiments. If $it=1$, the threshold  is equal to $\theta_{pr}$. As the number of iteration increases the  threshold is updated according to the product of $\alpha$ and iteration number $it$. This approach can make sure that we are picking samples with acceptable confidence. Algorithm~\ref{alg:UpDWSeT} presents our approach for updating labeled and unlabaled datasets.

\begin{algorithm}
	\caption{\small \textbf{Update Weighted Self-Training Datasets Algorithm (\textsc{UpDWSeT})}}\label{alg:UpDWSeT}
	{
		\begin{algorithmic}[1]
			\Procedure{UpDWSet}{$L_t, L_d, U, \alpha, \beta, \theta_{pr}, it$}
			\State $S=\emptyset$
			\For {$ l \in [True, False]$}
			\State {$thr = FindPrecisionThreshold (L_d, \theta_{pr}-\alpha \times (it-1) , label=l )$}
			\State {$S = S \cup \{\textbf{u} \in U|\textbf{u}.p\ge thr\}$} 
			\EndFor
			\State {$U= U - S$}
			\State {$S.w_c = \exp(-\beta \times it \times (\frac{1}{1-p}))$}
			\State{$L_t = L_t \cup S$}
			\State \textbf{return} $L_t, U$
			\EndProcedure
	\end{algorithmic}}
\end{algorithm}

\section{ISIS  Dataset}\label{sec:isisdata}
Our dataset consists of ISIS related tweets/retweets in Arabic gathered from Feb. 2016 to May 2016. This dataset is discussed in details in~\cite{shaabani2018detecting}. The dataset includes tweets and the associated information such as user ID, re-tweet ID, hashtags, number of followers, number of followee, content, date and time. About 53M tweets are collected based on the 290 hashtags such as State of the Islamic-Caliphate, and Islamic State. In this paper, we only use tweets (more than 9M) associated with viral cascades.

\begin{figure}
	\centering
		\includegraphics[width=.4\columnwidth]{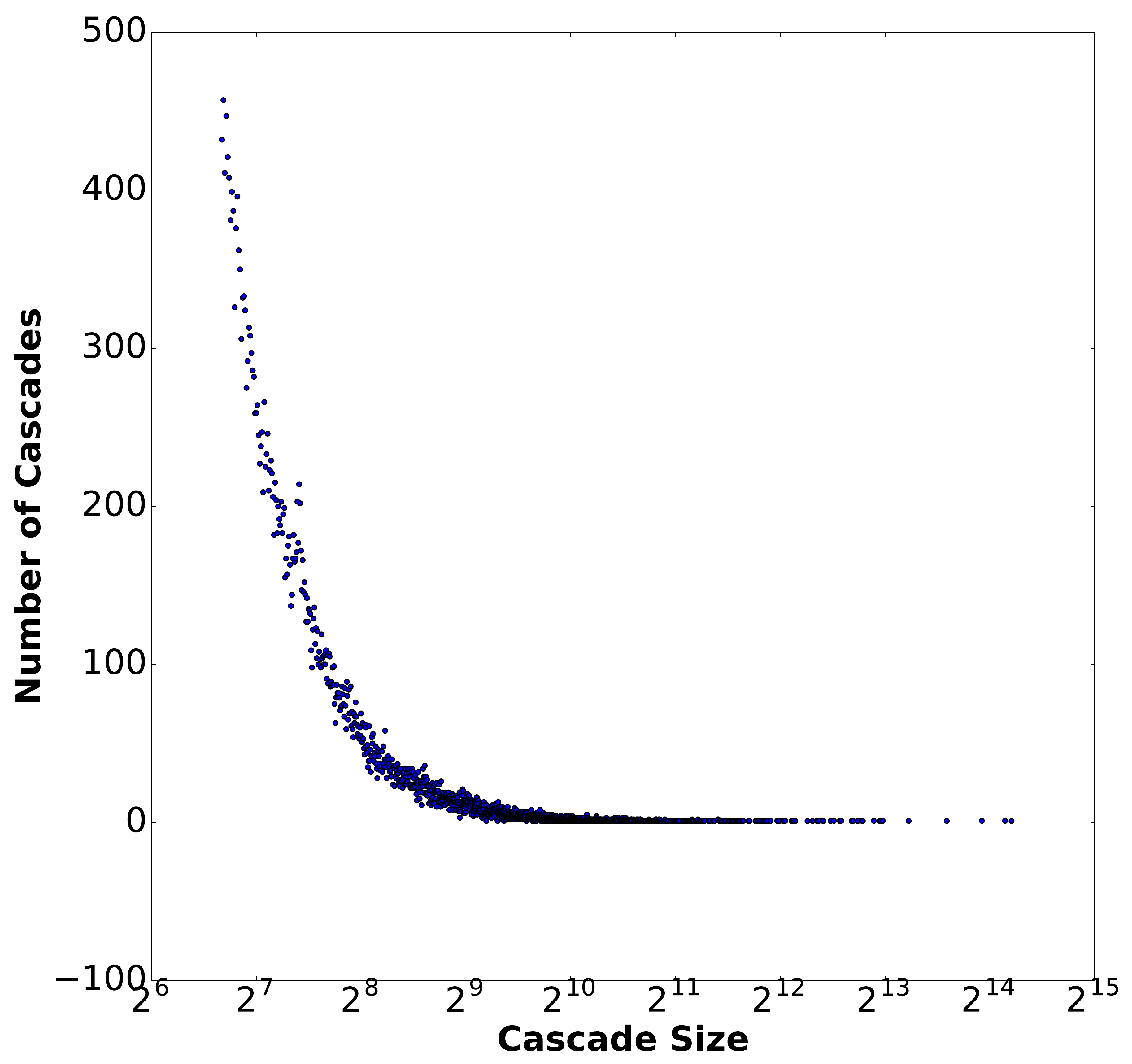}
	\caption{Distribution of cascades vs cascade size} 
	\label{fig:dist_casc_all}
\end{figure}

\noindent{\textbf{Cascades.}} We aim to identify PSM accounts.  For this dataset, they are mainly social bots or terrorism-supporting accounts that participate in viral cascades. The tweets  have been retweeted from 102 to 18,892 times. This leads to more than 35K cascades which are tweeted or retweeted by more than 1M users. The distribution of the number of cascades vs cascade size is illustrated in Fig.~\ref{fig:dist_casc_all}. 

\noindent{\textbf{User's Current Status.}} We select \textit{key users} that have tweeted or retweeted a post in its early life span - among first half of the users (according to Definition~\ref{def:kusers}, $\phi = 0.5$), and check whether they are active or not. Accounts are not active if they are suspended or deleted. Less than 24\% of the users are inactive. 
 Inactive users are representative of automatic and terrorism accounts aiming to disseminate their propaganda and manipulate the statistics of the hashtags of their interest. 
 
\section{Results and Discussion}\label{sec:res}
We implement part of our code in Scala Spark and Python 2.7x and run it on a machine equipped with an Intel Xeon CPU (2 processors of 2.4 GHz) with 256 GB of RAM running Windows 7. We also implement most of structural metrics in Cypher query language. We create the graphs using Neo4j~\cite{webber2018programmatic} on a machine equipped with an Intel Xeon CPU (2 processors of 2.4 GHz) with 520 GB of RAM. We set the parameter $\phi$ as 0.5 to label key users  (Definition~\ref{def:kusers}). That is, we are looking for the users that participate in the \emph{action} before the number of participants gets twice. 

In the following sections, first we look at the baseline methods. Then we address  the performance of two proposed approaches (see Section~\ref{sec:algo}): (1) \textit{Supervised Learning Approach}: applying different supervised learning methods on proposed metrics, (2) \textit{Self-Training Semi-Supervised Learning Approach}: selecting users by applying Algorithm~\ref{alg:WSeT}. The intuition behind this approach is to select users with the high probability of being either PSM or non-PSM (normal user) from unlabeled data and then adding them to the training set in order to improve the performance. We evaluate methods based on both Precision-Recall and Receiver Operating Characteristics (ROC) curves. Note that in all experiments, the training, development, and test sets are imbalanced with more  normal users than PSM users. The statistics of the datasets are presented in Table~\ref{tab:dataset_exp}. Dataset $\A$ is randomly selected from dataset $\B$ using sklearn library~\cite{scikit-learn}. Note that, all random selections of data in the experiments have been done using sklearn library. We repeated the experiments 3 times and picked the median output. It is worth to mention that the variance among the results was negligible. In this problem, our goal is to achieve high precision while maximizing the recall. The main reason is labeling an account as PSM means it should be deleted. However, removing a  normal user is costly. Therefore, it is important to have a high precision to prevent removing the normal users.

\begin{table}[ht]
	\centering
	\caption{\small \textmd{Statistics of the datasets used in experiments.}}
	{
		\begin{tabular}{|c | c|c|c|}
			\hline
			\textbf{Name} & \textbf{ PSM accounts} & \textbf{Normal accounts} & \textbf{Total}\\ \hline \hline
			 $\A$ & 19,859  & 65,417 & 85,276\\ \hline 
			$\B$ & 137,248 & 585,396 & 722,644\\ \hline			  
	\end{tabular}}
	\label{tab:dataset_exp}
\end{table}
 
\subsection{Baseline Methods}
We have compared our results with existing work for detecting PSM accounts~\cite{shaabani2018detecting,alvari2018early} or bots~\cite{subrahmanian2016darpa}.

\noindent\textbf{Causality.} This paper presents a set of causality metrics and unsupervised label propagation model to identify PSM accounts~\cite{shaabani2018detecting}. However, since our approach is supervised, we only use the causality metrics  and evaluate its performance in a supervised framework.

\noindent\textbf{C2DC.} This approach uses time decay causal community detection-based classification to detect PSM accounts~\cite{alvari2018early}. We also considered time decay causal metrics with random forest as another baseline method (TDCausality).

\noindent\textbf{Sentimetrix.} This approach is proposed by the top-ranked team in the DARPA Twitter Bot Challenge~\cite{subrahmanian2016darpa}. We consider all features that we could extract from our dataset. Our features include tweet syntax (average number of hashtags, average number of user mentions, average number of links, average number of special characters), tweet semantics (LDA topics), and user behaviour (tweet spread, tweet frequency, tweet repeats). The proposed method starts with a small seed set and propagate the labels. As we have enough labeled dataset for the training set, we use  random forest as the learning approach. 

\begin{figure}[t!]
	\centering
	\begin{subfigure}[b]{.5\columnwidth}
		\centering
		\includegraphics[width=1\columnwidth]{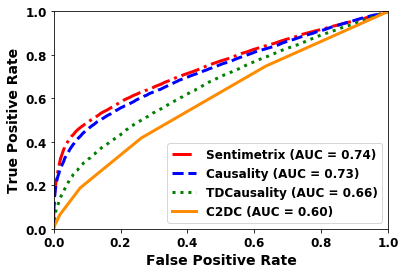}
		\caption{ROC curve} 
		\label{fig:sup_bl_roc}
	\end{subfigure}%
	\begin{subfigure}[b]{.5\columnwidth}
		\centering
		\includegraphics[width=1\columnwidth]{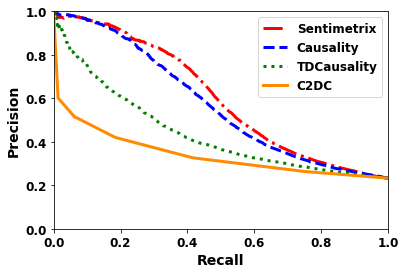}
		\caption{Precision-Recall curve}
		\label{fig:sup_bl_prrc}
	\end{subfigure}%
	\caption{Performance of the baseline methods on dataset $\A$} 
	\label{fig:dist_casc}	
\end{figure}

We use dataset $\A$ to evaluate different approaches. Fig.~\ref{fig:sup_bl_roc} shows the precision-recall curve for these methods. As it is shown, in the supervised framework, \textit{Sentimetrix} outperforms all approaches in general. Also, \textit{Causality} is a comparable approach with Sentimetrix with the constraint that the precision is no less than 0.9 as illustrated in Fig.~\ref{fig:sup_bl_prrc}.
Note that, most of the features used in the previous bot detection work take advantage of  content and network structure of  users. However, this is not the case in our proposed metrics and approach. 

\subsection{Supervised Learning Approach}
 In this section, we describe the classification results using proposed metrics with different learning approaches. We used both datasets for this experiment. First, we use the same dataset as we used in baseline experiments ($\A$). Then, we use dataset $\B$ for comparing top methods which is 8.5 times larger than dataset $\A$. 
 
\begin{figure}[t!]
	\centering
	\begin{subfigure}[b]{.5\columnwidth}
		\centering
		\includegraphics[width=1\columnwidth]{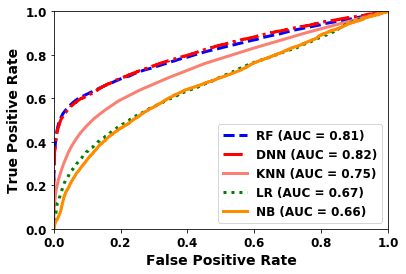}
		\caption{ROC curve} 
		\label{fig:sup_mt_roc}
	\end{subfigure}%
\begin{subfigure}[b]{.5\columnwidth}
		\centering
		\includegraphics[width=1\columnwidth]{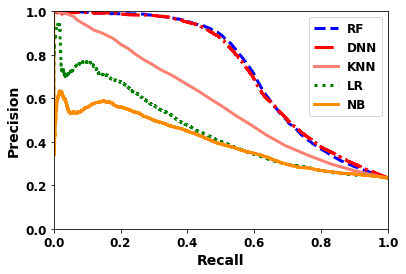}
		\caption{Precision-Recall curve}
		\label{fig:sup_mt_prrc}
	\end{subfigure}%
	\caption{Performance of different supervised approaches using proposed metrics on dataset $\A$} 
	\label{fig:dist_casc}	
\end{figure}

 Fig.~\ref{fig:sup_mt_roc} shows the ROC curve for different approaches. As it is shown, deep neural network achieved the highest area under the curve. Note that, the deep neural network is comparable with random forest as it is shown in Fig.~\ref{fig:sup_mt_prrc} on Dataset $\A$. The proposed approaches could improve the recall from 0.22 to 0.49 with the precision of 0.9. According to the random forest, top features are from all categories including causality metrics: $\epsilon_{nb}$, $\epsilon_{wnb}$, user-message graph-based metrics: user's PageRank $PR(u)$, median of retweeted message's PageRank $PS_{u,med}$, degree $D_u$, mean of verified users in his messages $U^vS_{u,mean}$, $\T S_{u,med}$, $\T S_{u,mean}$, median of length of the cascades $CS_{u,med}$, user graph-based metrics: weighted co-occurrence score $\CO^w_{u,N_u}$.

\begin{figure}[t!]
	\centering
	\begin{subfigure}[b]{.5\columnwidth}
		\centering
		\includegraphics[width=1\columnwidth]{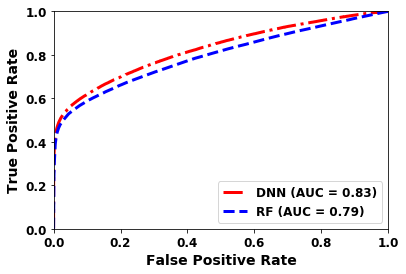}
		\caption{ROC curve} 
		\label{fig:sup_mt_2_roc}
	\end{subfigure}%
	\begin{subfigure}[b]{.5\columnwidth}
		\centering
		\includegraphics[width=1\columnwidth]{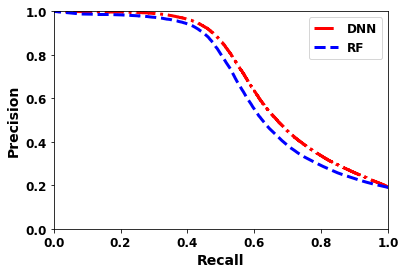}
		\caption{Precision-Recall curve}
		\label{fig:sup_mt_2_prrc}
	\end{subfigure}%
	\caption{Performance of the top two supervised approaches using proposed metrics on dataset $\B$} 
	\label{fig:dist_casc}	
\end{figure}

In Fig.~\ref{fig:sup_mt_2_prrc}, we probe the performance of the top two supervised approaches on larger dataset $\B$. Deep neural network is able to achieve the recall of 0.48 with the precision of 0.9. It is also able to achieve  AUC of 0.83 on this dataset (Fig.~\ref{fig:sup_mt_2_roc}). It is worth to mention that  we assigned higher weights to PSM accounts to deal with data imbalance problem.

\subsection{Self-Training Semi-Supervised Learning Approach}
\begin{figure}[t!]
	\centering
	\begin{subfigure}[b]{.5\columnwidth}
		\centering
		\includegraphics[width=1\columnwidth]{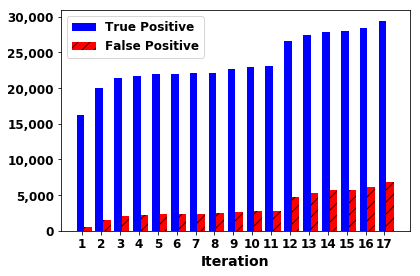}
		\caption{PSM users} 
		\label{fig:ss_pos}
	\end{subfigure}%
	\begin{subfigure}[b]{.5\columnwidth}
		\centering
		\includegraphics[width=1\columnwidth]{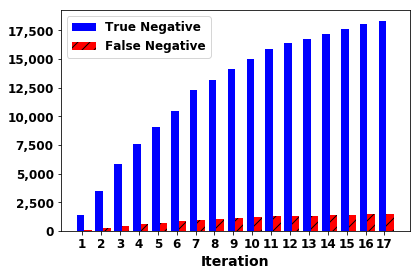}
		\caption{Normal users}
		\label{fig:ss_neg}
	\end{subfigure}%
	\caption{Cumulative number of selected users using \textsc{WSeT} Algorithm on dataset $\B$} 
	\label{fig:dist_casc}	
\end{figure}

In this experiment, we randomly select 300 PSM and 300  normal users from dataset $\B$ for training and development sets and the rest of the dataset was considered as unlabeled data. We conduct two types of experiments: 

\noindent\textbf{\textsc{WSeT} Algorithm.} In this experiment, we evaluate the self-training semi-supervised approach using Algorithm~\ref{alg:WSeT}. In this approach, we iteratively update the training set and the termination condition is accuracy of the model on the development set. We set the parameters as $\theta_{pr}=1$, $\alpha=0.05$, $\theta_{tr}=0.03$. We use a random forest classifier to train the model. The cumulative number of true positive and false positive is shown in Fig.~\ref{fig:ss_pos}. With using 300 PSM accounts as seed set, \textsc{WSeT} can find 29,440 PSM accounts with the precision of 0.81. Note that, we can stop the algorithm earlier. 

In this case, precision varies from $0.97$ to $0.81$. Fig.~\ref{fig:ss_neg} illustrates cumulative number of selected users as normal users by \textsc{WSeT}. As shown, the number of true negatives (selected normal accounts as normal users) is 18,343 with precision of 0.93. 

\noindent\textbf{Supervised \textsc{WSeT} Algorithm.} In previous experiment, we assume that the supervisor checks accounts labeled as PSM by \textsc{WSeT} at the end. However, this process can be done iteratively. Here, we assume that the supervisor evaluates the \textit{PSM labeled accounts by \textsc{WSeT}} in each iteration and verify if they are either true or false positive. Therefore, these labels along with the non-PSM labeled accounts by \textsc{WSeT} are fed into \textsc{WSeT}. According to our results, the number of true positive increases to 80,652 with the precision of more than 0.8. That is, using this approach we can increase the number of true positive PSM accounts 2.7 times.

\begin{figure}
	\centering
	\includegraphics[width=.7\columnwidth]{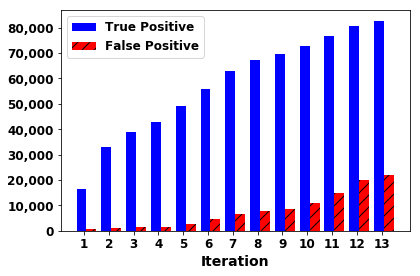}
	\caption{Cumulative number of selected users as PSM accounts using supervised \textsc{WSeT} Algorithm on dataset $\B$} 
	\label{fig:ss_act}	
\end{figure}

\section{Related Work}\label{sec:rw}

In summary, majority of previous work was based on three fundamental assumptions which make them different from our work. First, \emph{the information of the network is known}~\cite{subrahmanian2016darpa,goyal2010learning,benignitweets,abokhodair2015dissecting}. This assumption may not hold in reality. Second, \emph{they are language-dependent} ~\cite{subrahmanian2016darpa,morstatter2016new}. Third, \emph{the majority of botnet detection algorithms focused on bots in general.} That is, they did only consider the bots \emph{equally}~\cite{morstatter2016new,dickerson2014using} while here, we identify PSM accounts that spread harmful viral information. 

\noindent\textbf{Identifying PSM accounts.} Compared to previous PSM account detection work in \cite{shaabani2018detecting, alvari2018early}, we propose graph-based features and unlike the unsupervised learning approach in \cite{shaabani2018detecting}, we expand out both the supervised and semi-supervised learning methods. We also develop a deep neural network  and show  how our proposed approaches improve them significantly.

\noindent\textbf{Identifying Automatic Accounts.}

DARPA conducted the Twitter bot detection challenge to identify and eliminate influential bots~\cite{subrahmanian2016darpa}. Most of the previous work extracted different sets of features (tweet syntax, tweet semantics, temporal behavior, user profile, friends and network features) and conducted supervised or semi-supervised approaches~\cite{subrahmanian2016darpa,davis2016botornot,dickerson2014using}. However,  without using content, and network structure, they perform poorly. Also, some of the features such as tweet semantics depend on the language. It is yet a challenge to apply these features to other languages such as Arabic.  

\noindent{\textbf{Analysis of Terrorist Groups and Detection of Water Armies.}} Terrorist groups use social media for propaganda dissemination~\cite{al2015examining}. Benigni et al.~\cite{benignitweets} conducted vertex clustering and classification to find Islamic Jihad Supporting Community on Twitter. Abdokhodair et al. \cite{abokhodair2015dissecting} studied the behaviors and characteristics of Syrian social botnet. Chen et al.~\cite{chen2013battling} found that within the context of news report comments, user-specific measurements can distinguish water army from normal users. Our work is different from them since these methods  used features related to the accounts and network. 

\section{Conclusion}
In this paper, we conducted a data-driven study on the pathogenic social media accounts. We proposed supervised and semi-supervised frameworks to detect these users. We achieved the precision of 0.9 with F1 score of 0.63 using supervised framework. In semi-supervised framework, we are able to detect more than 29K PSM users by using only 600 labeled data for training and development sets with the precision of 0.81. Our approaches identify these users without using network structure, cascade path information, content and users' information. We believe our technique can be applied in the areas such as detection of water armies and fake news campaigns. Our future plan is to combine the proposed semi-supervised approach with unsupervised one  in~\cite{shaabani2018detecting} in order to find the initial seed set for the semi-supervised approach.\medskip

{\small
	\noindent\textbf{Acknowledgments.} Some of the authors are supported through the ARO (grant  W911NF-15-1-0282).
	
\bibliographystyle{IEEEtran}
{

}
\end{document}